\documentstyle[psfig]{ioplppt}
 \begin{document}

 \jl{4}
 \title{High energy hadrons in EAS at mountain altitude}
  [High energy hadrons in EAS]
 \author{J~N~Capdevielle\dag, J~Gawin\ddag, D~Sobczy\'{n}ska\S,
 B~Szabelska\ddag, J~Szabelski\ddag, and T~Wibig\S}
 
 \address{\dag\ Laboratoire de Physique Corpusculaire, Coll\`ege de France, 
 11 place Marcelin Berthelot, 75231 Paris 
 Cedex 05, France}
  
 \address{\ddag\ The Andrzej So{\l }tan Institute for Nuclear Studies, 
 90-950 {\L }\'{o}d\'{z}, Box 447, Poland}
 
 \address{\S\ University of {\L }\'{o}d\'{z}, Experimental Physics Dept.,
 ul.~Pomorska 149/153, 90-236 {\L }\'{o}d\'{z}, Poland}

 \begin{abstract}

 An extensive simulation has been carried out to estimate
 the physical interpretation of dynamical factors such as
 $< R_{h} >$, $< E_{h} R_{h} >$ in terms of high energy interaction 
 features, concentrated in the
 present analysis on the average transverse momentum.\\
 It appears that the large enhancement observed for
 $< E_{h} R_{h} >$ versus primary energy, suggesting in earliest
 analysis a significant rise of $< p_{t} >$ with energy, is only
 the result of the limited resolution of the detectors
 and remains in agreement with a wide range of models used
 in simulations.

 \end{abstract}
 \maketitle

 \section{Introduction}

 Interest in the search for additional information in the hadron 
 component of EAS at mountain altitude in terms of very high
 energy interactions \cite{Nikolsky} has been spurred by the necessity
 to have a continuity of multiproduction features
 between present colliders at $\sqrt{s} \le$~1.8~TeV and the 
 future LHC high energy physics at $\sqrt{s}$~=~16~TeV
 ($\sqrt{s}$ represents here the centre of mass energy
 for the nucleon--nucleon collision).

 We review here the data obtained from calorimeters, or
 equivalent detectors, on high energy hadrons (longitudinal
 development and lateral spread) coupled with air shower
 arrays, where the primary energy is estimated from the
 electron and the muon components.

 From qualitative considerations, the most energetic hadrons
 ($E_{h} > 1~TeV$) can be supposed to conserve some significant
 properties correlated with the dynamics of the earliest 
 interactions: for instance, the relation between the transverse
 momentum $p_{t}$, the height h of the interaction, the hadron
 energy $E_{h}$ and the radial distance $R_{h}$,
 
 \begin{equation}
        p_{t} = \frac{R_{h}}{h} \, \frac{E_{h}}{c}
 \end{equation}
 \noindent
 has been suggested as a possible measure of $p_{t}$ \cite{Winn}.\\
 
 Following this line of inquiry, special attention has been given 
 to the measurements of quantities like $<E_{h} R_{h} >$, $<E_{h} >$,
 $< R_{h} >$ (where $<>$ are the mean values per hadron -- see
 \cite{Romakhin77, NestRom})
 and $n_{h}$, the number of hadrons above the energy
 threshold $E_{th}$. A convergent and significant increase of
 $< E_{h} R_{h} >$ was observed in the different experiments
 \cite{Aseikin, Sreekantan, Bohm}
 and the earliest interpretation was a tremendous
 rise of the transverse momentum $<p_{t}>$, above $\sqrt{s}$~=~5000~GeV,
 the minimum for $p_{t}$ (occuring from equation 1 when the hadrons 
 come directly from the first interaction) being situated around
 2~GeV/c \cite{Winn}.\\
 In more recent analysis measurements of hadron lateral distribution
 in Tien Shan calorimeter
 (which unfortunately have never been repeated
 \cite{NRprivate}) have been interpreted in terms of increasing role
 of high $p_{t}$ processes \cite{JadFiz}.
 The problem still remains open, since the number of reports on the 
 subjest is small and \lq\lq measurements of hadrons suffer more than 
 other EAS components from systematic errors which have to be carefully 
 estimated, using simulations of the detector response" \cite{Sivaprasad}.
 
 In parallel with our review of the experimental data, we have
 carried out EAS simulations with the program CORSIKA version 4.50
 \cite{Knapp, Capd92},
 regarded as containing the major features of the present collider physics,
 in the energy range 10$^{5}$ -- 10$^{8}$~GeV.

 Our analysis has concentrated on the hadron lateral spread
 to explain the specific effects on the different factors combined
 in $<E_{h} R_{h} >$, such as the sum of energy deposited, 
 $\Sigma{E_{h}}$, the hadron distance, $R_{h}$, and the hadron content
 $n_{h}$ (with consequences for the containment as well as the resolution
 of the detector). In order to appreciate the complexity of the
 hadron cascade mechanism, and how it can conserve or smear out
 the original interaction characteristics, the progression of
 our analysis is organized as follows:
 \begin{itemize}
 \item physical dependence of interaction parameters (multiplicity,
       inelasticity, $p_{t}$ distribution....)
 \item dependence on primary spectrum and composition
 \item dependence on data acquisition and experimental resolution.
 \end{itemize}

 The survey of this dependence has  focused on the largest 
 detector employed, i.e. the Tien Shan calorimeter of 36~m$^{2}$ 
 \cite{Aseikin}.

 \section{Simulation of the hadron lateral spread.}
 \subsection{Assumptions on $p_{t}$ and CORSIKA options.}

\begin{figure}[h]
 \begin{center}
 \mbox{
 \psfig{file=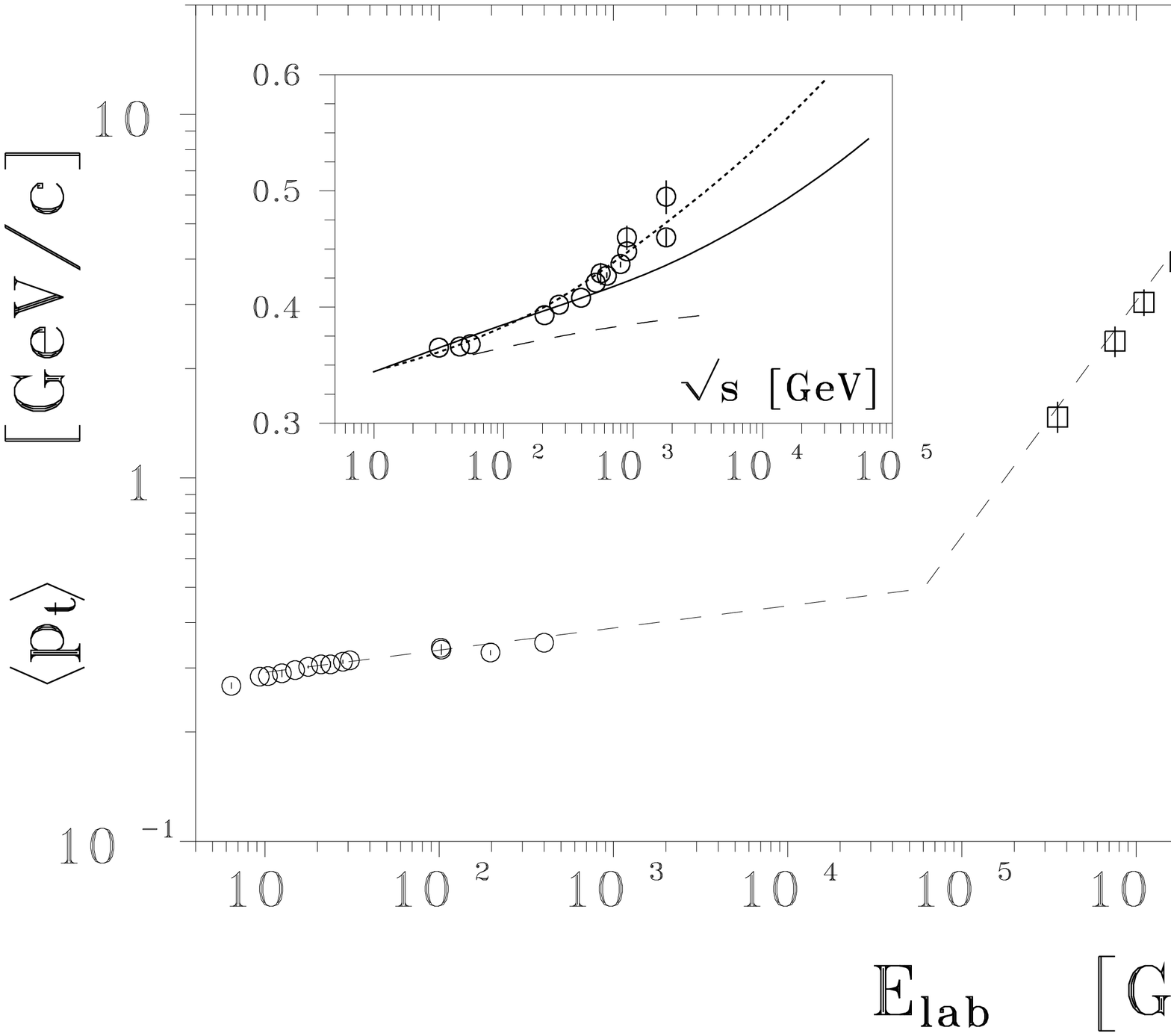,width=8cm}
 }\\
 \end{center}
 \caption[mean pt versus primary energy]{Dependence
 of $<p_{t}>$ versus primary energy with the data interpretation
 of ref. \cite{Winn} from $<E_{h}R_{h}>$. In the window options
 in CORSIKA adapted from ref. \cite{Knapp}:\\
  -- 1. HDPM + CERN UA1 85 (solid line)\\
  -- 2. HDPM + CERN UA1 Mimi 96 (dotted line)\\
  -- 3. VENUS (dashed line).\\
  $\opencirc$     accelerator and collider data compiled in \cite{Knapp}\\
  $\opensqr \, \,    <E_{h}R_{h}>$ interpretation
        from the Tien Shan experiment in \cite{Winn}.}
 \label{fig:ptve}
 \end{figure}

 The angular emission of secondary hadrons depends on
 the ratio 
 $p_{t}/p_{l} \sim p_{t}/E_{h}$, 
 where $p_{l}$ is 
 longitudinal momentum in laboratory system.
 Its evolution
 from CM frame to laboratory system
 is dominated, after Lorentz transformation,
 by the well known \lq\lq collimation effect", the emitted jet
 becoming narrower as the hadron energy increases. 
 The multiplicity grows as $\ln^{2}{s}$, inelasticity 
 changes slowly and $p_{t}$ rises regularly \cite{Knapp}, but
 in reasonable proportion (circle shape points in figure
 ~\ref{fig:ptve}).
 
 Other effects, such as the increase in $p_{t}$ for p--A collisions or A--A
 collisions (as described for the \lq\lq Cronin effect"
 by Schmidt and Schukraft \cite{Schmidt}), or correlated with jet production,
 or the increase in $p_{t}$ for diffractive events \cite{Goulianos} are not 
 very important or can be disregarded according to their
 respective cross sections.

 The correlation between $<p_{t}>$ and
 multiplicity (or central rapidity density $dn/dy$) 
 cannot be
 neglected in the treatment of individual events more correlated
 with the semi--inclusive data; the correlation of the UA1 experiment,
 described in \cite{Schmidt}, is included in CORSIKA package (for the option
 HDPM) and has been used, without change, in the main part of 
 our simulation. 
 We have, nevertheless, performed some simulations
 with a $p_{t}$ generator following the most recent correlation
 $<p_{t}>$ with $dn/dy$ 
 from the CERN UA1 MIMI collaboration \cite{Bocquet},
 characterized
 by a greater rise of $<p_{t}>$ at large multiplicity.
 In this way, we
 ascertained that the subsequent relative changes in $<E_{h}R_{h}>$
 are of second order when compared 
 to the attempt 
 to explain $<E_{h}R_{h}>$ with $<p_{t}>$ values exceeding
 2~GeV/c at $\sqrt{s}$~=~5~TeV \cite{Winn}
 (squares on figure~\ref{fig:ptve}).
 
 \begin{figure}[h]
 \begin{center}
 \mbox{
 \psfig{file=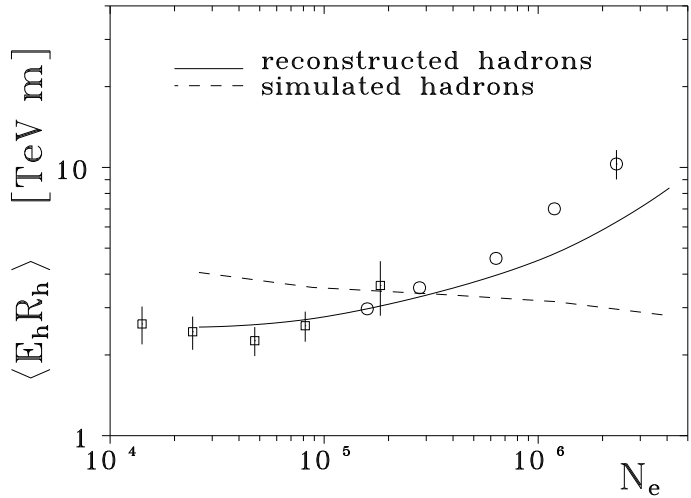,width=8cm}
 }\\
 \end{center}
 \caption[...]{
 Dependence of $<E_{h}R_{h}>$ versus $N_{e}$ 
 (for $E_{th}$~=1~TeV).\\
   --  exact simulation (HDPM) output (dashed line)\\
   --  after reconstruction described in section 3 (solid line)\\
  $\opensqr$   Tien Shan data \cite{Machavariani}\\
  $\opencirc$  Tien Shan data \cite{RomNest}}
 \label{fig:ervne}
 \end{figure}
 
 Furthermore some simulations have been done with this extremal value \\
 $<p_{t}>$~=~2~GeV/c around a primary energy $E_{0}$~=~10$^{7}$~GeV,
 after a special tuning of the $p_{t}$ generator in CORSIKA:
 the options currently used have been HDPM in the usual form \cite{Capd92},
 as well as coupled to the $p_{t}$ generator elaborated following
 UA1 MIMI results \cite{Capd97}, and also the VENUS option \cite{Capd92}. 
 A detailed
 description of the evolution with energy of the global interaction
 parameters used in the options of CORSIKA can be found in 
 references \cite{Knapp} -- \cite{Bocquet}.\\
    
 The physical area explored following our selection of input parameters
 concerns primary energies between 10$^{5}$~--~10$^{8}$~GeV, 
 zenith angles between 0 and 30 deg for primary masses A = 1, 4, 14, 28, 56. 
 The altitude for observation levels has been set to 3200~m 
 with statistical samples of 500 simulated cascades per primary.
 Showers induced by heavy primaries have been simulated simultaneously with
 VENUS option and HDPM combined with superposition or , alternatively, with
 the abrasion-evaporation procedure \cite{Attallah}, describing the 
 fragmentation of heavy projectiles.
    
 \subsection{Consequences for the hadron lateral spread.}
 
 The results obtained for 
 $<E_{h}R_{h}> = (\Sigma{E_{h}R_{h}})/n_{h}$
 are shown in the  figure~\ref{fig:ervne} (dashed line), 
 where $<E_{h}R_{h}>$ is plotted
 versus electron size $N_{e}$, for proton primaries and for a
 hadron energy threshold $E_{th}$~=~1~TeV. The expected decrease
 versus size, as in previous calculations \cite{NestRom}, is obtained.
 The agreement is tolerable only for the experimental data
 \cite{Machavariani} at $N_{e}$ near $2 \cdot 10^{5}$.
 The lateral hadron densities $\Delta_{h}$
 have been described by the experimenters with a simple
 exponential law:
 \begin{equation}
         \Delta_{h} = \frac{n_{h}}{2 \pi R_{0}^{2}} \, exp{(-\frac{R_{h}}{R_{0}})}.
 \end{equation} 
 
\begin{figure}[b]
 \begin{center}
 \mbox{
 \psfig{file=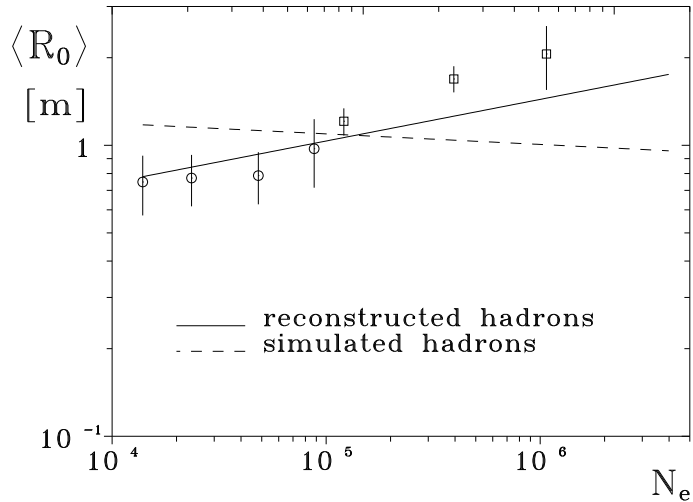,width=8cm}
 }\\
 \end{center}
 \caption[...]{ 
 Dependence of $<R_{0}>$ versus $N_{e}$. (Lines and experimental data
 as in figure~\ref{fig:ervne}).}
 \label{fig:rvne}
 \end{figure}

 \noindent
 We have used the same functional dependence to fit the hadron 
 distribution in each individual shower with two free parameters: 
 normalization related to the total number of hadrons $n_h$ 
 and mean radius $R_0$. 
 The total number of hadrons $ <n_h> $ varied from 0.05 for HDPM model
 and primary Fe
 to 0.9 for VENUS option and primary proton for $N_{e}$ = 10$^{5}$
 and respectively from 3.5 to 9.0 for $N_{e}$ = 10$^{6}$.
 The average mean radius as a function 
 of the shower size is presented in figure~\ref{fig:rvne}.
 Similar remarks as in  figure~\ref{fig:ervne} are suggested by  
 figure~\ref{fig:rvne}, where
 $R_{0}$ decreases versus size, opposite to the general
 experimental tendency.
 
 \newpage
 \begin{table}
 \caption[....]{
 Dependence of $<E_{h}R_{h}>$ and $<R_{0}>$ on interaction
 and shower parameters 
 from the simulations as noted.
 }
 \lineup
 \begin{indented}
 \item[]\begin{tabular}{@{}lllll} 
 \br
   Factors              & \0A  &$<E_{h}R_{h}>$ & \0\0$<R_{0}>$   & comment       \\
                        &      & (TeV m)       & \0\0(m)    &                  \\
 \mr
 a) primary mass A       &\01  &\0\03.16 & \0\01.6    & HDPM +           \\
 at $E_{0}=10^{6}$GeV    &\04  &\0\03.4  & \0\01.70   & abrasion-        \\
                         & 14  &\0\03.67 & \0\02.21   & evaporation \cite{Attallah} \\
                         & 28  &\0\04.29 & \0\02.52   & + UA1 MIMI     \\
                         & 56  &\0\04.7  & \0\03.24   & 96 \cite{Bocquet} \\
                         &     &         &            &                \\
                         & 56  &\0\03.81 & \0\01.34   & HDPM+superp.   \\
                         &     &         &            &                \\
                         &\01  &\0\03.73 & \0\01.09   & Venus          \\
                         &     &         &            &                \\
 \br                         
                         &\0$E_{th}$ &$<E_{h}R_{h}>$ & \0\0$<R_{0}>$   & comment \\
                         & (TeV)     & (TeV m)       & \0\0(m)       &         \\
 \mr
 b) hadron               & \00.32    & \0\01.988     & \0\01.24      &  HDPM \\
 threshold energy $E_{th}$& \01.0    & \0\03.34      & \0\01.01      &       \\
 (primary protons        & \03.16    & \0\05.51      & \0\00.765     &       \\
 of $10^{6}$GeV)         & 10.0      & \0\08.29      & \0\00.57      &       \\
                         &           &               &               &       \\
 \br                         
                         &\0\0$t_{0}$ & \0\0\0z &$<E_{h}R_{h}>$ & \0$<R_{0}>$  \\
                         &($gcm^{-2}$)&         &  (TeV m)      & \0\0(m)      \\
 \mr
 c) initial conditions,  &  \0\05  & \0\0\02  & \0\05.3   &   \0\03.08    \\
 multiplicity            &  \0\05  & \0\01/2  & \0\05.6   &   \0\02.78    \\
 ($z=n_{ch}/<n_{ch}>$)   &   150   & \0\0\02  & \0\02.4   &   \0\00.98    \\
 and depth ($t_{0}$) of first& 150 & \0\01/2  & \0\02.5   &   \0\01.26    \\
 interaction (primary    &         &          &           &             \\
 protons of $10^{6}$GeV) &         & option:  & HDPM      &             \\
 \br                         
                         & \0$E_{0}$& $<E_{h}R_{h}>$  & \0\0\0$<R_{0}>$ &  option \\
                         & (GeV)    & (TeV m)         & \0\0\0(m)     &         \\
 \mr
 d) primary energy       & \0~10$^{5}$ & \0\04.248      &  \0\01.122 &  HDPM \\
 (for protons        &3$\cdot$10$^{5}$ & \0\03.678      &  \0\01.08  &       \\
 and for $E_{th}$~=~1~TeV)& \0~10$^{6}$ & \0\03.58       &  \0\01.03  &       \\
                     &3$\cdot$10$^{6}$ & \0\03.34       &  \0\01.01  &       \\
                         & \0~10$^{7}$ & \0\03.00       &  \0\00.95  &       \\
                         & \0~10$^{8}$ & \0\02.69       &  \0\00.93  &       \\
 \br                     
 \end{tabular}
 \end{indented}
 \end{table}
 
 The total number of hadrons $n_{h}$ reflects also a disagreement,
 characterized by an excess in the simulation \cite{Dubovy},
 increasing with primary energy: the amplitude of this
 disagreement, when compared to the influence of primary energy
 and interaction parameters, is illustrated by the values calculated
 in table  1.\\
 
 Setting $<p_{t}>$ above 2~GeV/c around a primary energy 
 $E_{0}$~=~10$^{7}$~GeV gave no rise in $<E_{h}R_{h}>$
 and $<R_{0}>$: a slight decrease was even observed due to the larger 
 transverse mass per secondary and to the consequent enhancement
 of the inelasticity in the former collision, reducing in proportion
 $\Sigma E_{h}$ at mountain level.\\
 
 Table 1 represents a small part of the total simulation which 
 contains similar effects to the examples presented. The total
 sample, up to 10$^{17}$~eV, confirms general aspects of the dependence,
 that we can summarize as follows:\\
 \begin{itemize}
 \item $<R_{0}>$ and $<E_{h}R_{h}>$ decrease with energy even if
       $<p_{t}>$ is rising with energy.
 \item This effect persists even for extremal value of 
       $<p_{t}>$~=~2~GeV/c and the \lq\lq collimation effect" dominates
       over all the energy interval considered.
 \item The height of the first interaction has an effect on the lateral
       spread, increasing $<E_{h}R_{h}>$.
 \item The factors $<R_{0}>$ and $<E_{h}R_{h}>$ increase with the
       primary mass A.
 \end{itemize}

 \section{Event topology and detector response.}

 The conditions of reception of the main hadron detectors
 employed at mountain altitude are compared in table 2.\\
 
 \begin{table}
 \caption[...]{Hadron detectors at mountain altitude
 \cite{Aseikin, Sreekantan, Bohm, Miyake, Matano, Ticona}.}
 \lineup
 \footnotesize\rm
 \begin{tabular}{@{}lllllll}
 \br
 Location     & Detector    & Area  &Energy range&Resolution& \0Altitude& Comment \\
              &           &($m^{2}$)& \0\0\0(TeV)& \0\0(cm )&(km a.s.l.)&         \\
 \mr
 Tien Shan    & ionization  & \036  & \0$\ge$ 0.3& \0\0\025 & \0\03.2  & axis \\
              & calorimeter &       &            &          &          & contained \\
 Chacaltaya I &em. chamber +& \0\08 &            & \0\0\050 & \0\05.2  &      \\
              & burst det.  &       &            &          &          &      \\
 Chacaltaya II&    "        & \0\08 & \0$ >$ 0.01& \0\0\050 & \0\05.2  &      \\
 Norikura     & proportional& \025  & 0.02 - 0.12& \0\0\050 & \0\02.77 & $20m<r$\\
              & counters +  &       &            &          &          & $r<150m$  \\
              & water tank +&       &            &          &          &      \\
              & lead plates &       &            &          &          &      \\
 \br
 \end{tabular}
 \end{table}
 
 The hadron energy in the Tien Shan ionization calorimeter, composed
 of 850~g~cm$^{-2}$ lead, is measured at a depth between 
 130 -- 850~g~cm$^{-2}$ with an accuracy of 20$\%$ 
 \cite{NestDub}.
 This thickness represents only 3.7 $\Lambda_{I}$ (for an interaction 
 length in lead $\Lambda_{I}$~=~194~g~cm$^{-2}$) corresponding to 
 a containment of about 80$\%$ of the cascades induced by individual 
 hadrons \cite{Bintinger}. The coordinates of the single hadrons
 are measured with an accuracy of 25~cm in the 744 ionization chambers
 of the calorimeter: those chambers, with an area of 
 (25 $\times$ 300)~cm$^{2}$, originally arranged in 15 rows, are intended 
 to measure the hadron number with an accuracy of 5$\%$
 \cite{Romakhin77}. Correction coefficients taking
 into account the limited lateral resolution and the possibility of
 aggregation of several hadrons within the chamber size bin
 (0.25 $\times$ 3)~m were introduced by Danilova \cite{Danilova}
 to recalculate the hadron energy spectra. The original method
 to separate and count the hadrons was based on the histograms
 of the ionization energy distribution in different rows in X and Y
 \cite{RomNest}. 
 \begin{figure}
 \mbox{
 \psfig{file=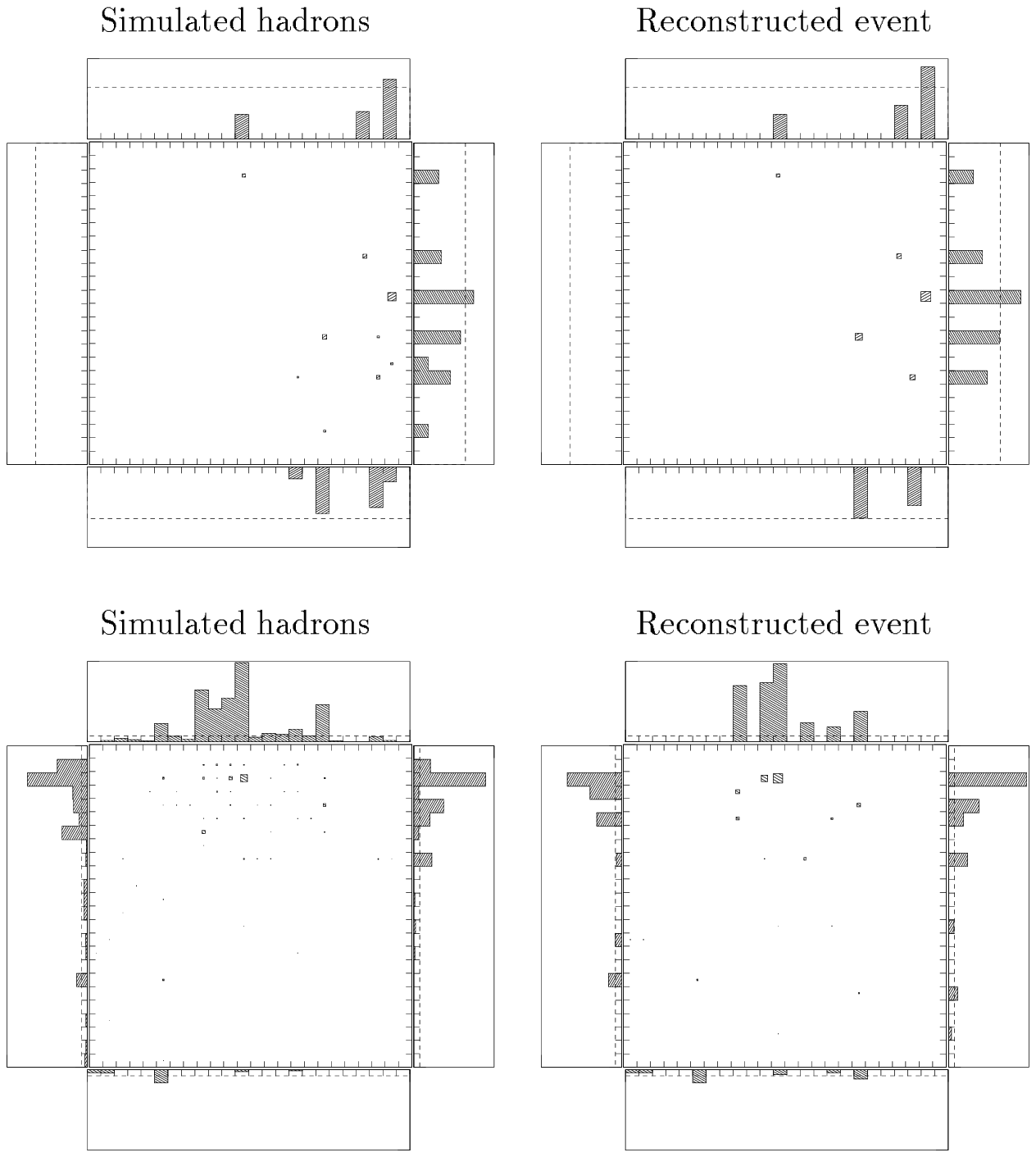,width=12cm}
 }\\
 \caption[...]{ 
 Two examples of total hadronic energy content of 10$^{6}$
 and 10$^{7}$~GeV proton showers (upper and lower plots respectively)
 in each 25~cm $\times$ 25~cm detector bin (and its 
 four one-dimensional
 projections as side histograms) 
 and the same after reconstruction procedure. 
 Dashed lines in projection plots show the 1~TeV hadron energy threshold
 and the discrepancy can be due to the reduction factor (by 2.6)
 used in Tien Shan experiment in case of hadrons superposition 
 \cite{Aseikin,RomNest}. See text for more detailed explanations 
 and comments.}
 \label{fig:exam}
 \end{figure}
 The limited efficiency of this 
 treatment can be seen in figure~\ref{fig:exam}, 
 which shows 
 real contained target diagrams for hadrons above 1~TeV
 and the same events reconstructed with the Tien 
 Shan procedure;
 other difficulties appear also in the geometrical containment,
 the fraction of hadrons registered in one shower decreasing
 strongly with the primary energy (from 60$\%$ at $N_{e}=10^{5}$
 down to 20$\%$ at $N_{e}=10^{6}$ according to our simulation).\\
 
\begin{figure}[h]
 \begin{center}
 \mbox{
 \psfig{file=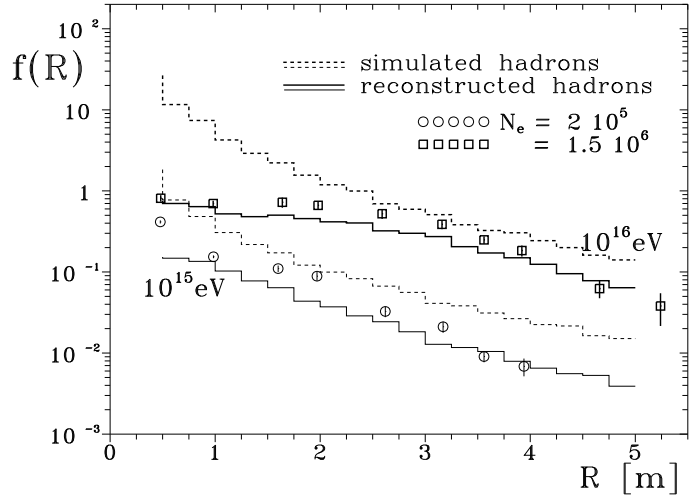,width=8cm}
 }\\
 \end{center}
 \caption[..]{ 
 Hadron lateral distribution function
 f(R) = dn$_{h}$/dR (for $E_{h} > 1 TeV$)
 simulated and recalculated
 at 10$^{6}$ and 10$^{7}$~GeV. Experimental data \cite{Danilova}.}
 \label{fig:latdis}
 \end{figure}

 The events simulated for section 2 have been treated
 by the numerical procedure of reconstruction of experimentally 
 recorded detector response.
 The energy of each hadron was added to a respective ionization chamber 
 output. As it is shown in figure~\ref{fig:exam} from a detector 
 plane four one-dimensional
 projections are obtained (each chamber is 3m long while the
 detector size is 6m $\times$ 6m). Then on each projection the possible 
 \lq\lq hadronic cluster" coordinates have been determined. 
 Using as a separation criterion the dip in the projected energy deposit 
 distribution of at least 70\% of 
 neighbouring
 chamber outputs, close chamber signals could be combined just 
 as it has been performed in the Tien Shan experiment.
 Then after comparing the amplitudes on respective x- and y-projections
 taking into account the possibility of overlapping of some hadrons in one
 of the projections the so-called \lq\lq hadronic clusters" have been found.
 They will be called hereafter the reconstructed hadrons. They are shown
 in the right plots in figure~\ref{fig:exam}.\\
 \begin{figure}[b]
 \begin{center}
 \mbox{
 \psfig{file=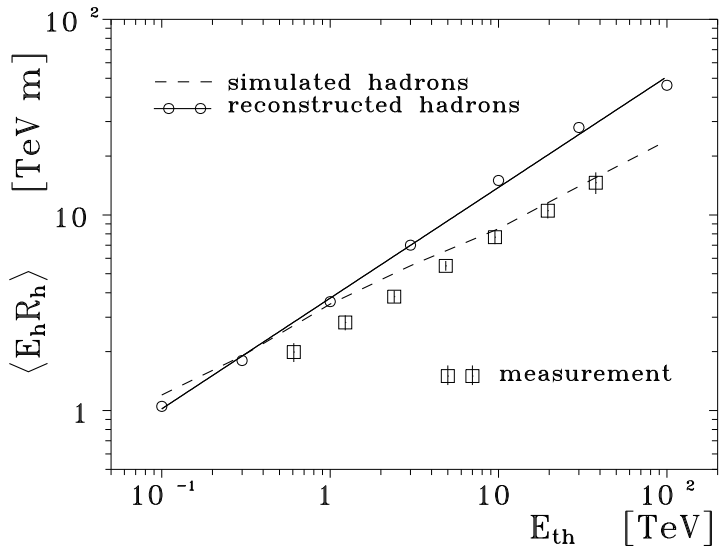,width=12cm}
 }\\
 \end{center}
 \caption[..]{
 Dependence of $<E_{h}R_{h}>$ on the energy threshold $E_{th}$.
 The squares are experimental points from \cite{Bohm} and \cite{RomNest}. 
 The energy dependence of the simulation
 is parallel to the experimental behaviour.}
 \label{fig:ervet}
 \end{figure}
 Our reconstruction procedure includes also hadrons of energies below the
 threshold (1TeV) which are mostly added to the large hadronic clusters.
 The correction coefficient of 2.64 (equal exactly to the one used by the
 Tien Shan group) have been applied to the cluster energy while 
 converted to the reconstructed hadron energy. The position of the 
 reconstructed hadron was the energy weighted mean of chamber outputs
 in each projection. 
 As the shower axis coordinates we have used the most energetic 
 reconstructed hadron coordinates just like it has been done in 
 Tien Shan experiment.\\
 For all situations considered in section 2 the general decrease
 for $<E_{h}R_{h}>$ and $<R_{h}>$ vs size turns to a large
 increase (solid line in figure~\ref{fig:ervne} and figure~\ref{fig:rvne}), 
 as well as $<E_{h}R_{h}>$ vs $E_{th}$
 (figure~\ref{fig:ervet}). This is confirmed (figure~\ref{fig:latdis}) 
 by comparison with the most
 recent data \cite{Danilova} where the individual reconstruction indicates
 how the hadron content is underestimated near the axis.
 To preserve the clarity of the graphs, we have not plotted
 on figures \ref{fig:ervne} and \ref{fig:rvne} the results of 
 the experiments 
 \cite{Sreekantan}, \cite{Bohm}, \cite{Miyake} -- \cite{Ticona}
 confirming the tendencies shown. Those values normalized to
 Tien Shan data can be found in reference \cite{Winn} and 
 reference \cite{RomNest}.

 \section{Discussion}
 
 The proportion between the factor $<R_{0}>$ calculated in
 table 1 and the resolution distance quoted in table 2
 shows the role of underestimation of hadron densities near the
 axis (when several hadrons enter simultaneously in the same 
 chamber bin). This effect will be even greater than in the Tien Shan
 ionization calorimeter in all the other detectors
 quoted . The values
 of $E_{th}$ correspond to larger hadron densities near the axis,
 increasing the probability of aggregation in the same chamber 
 bin. Also for larger resolution distance (i.e. wider bins)
 this probability increases.\\

 The simulation suggests that the hadron component and its lateral
 spread needs to be measured with more accurate calorimeters  
 and a resolution distance substantially smaller than 25~cm.\\

 We have shown that the rise $<E_{h}R_{h}>$  
 is a consequence only
 of the complex treatment of the data which is inadequate for the superposition
 of hadrons, 
 and the rise of $<p_{t}>$ derived from $<E_{h}R_{h}>$ and shown 
 in the figure~\ref{fig:ptve} is not valid. 
 The lateral spread of muons above 200 and 500~GeV,
 also correlated with $<p_{t}>$ in earliest interactions, doesn't
 require such an enhancement of $<p_{t}>$ \cite{Vashkevitch};
 the main bulk of these lateral muon distributions being 
 explained with a constant value of $<p_{t}>$~=~0.4~GeV/c
 \cite{Khrenov}.
 
 As follows from our simulations,
 even if $<p_{t}>$ is rising with energy, the hints in $<R_{0}>$
 and $<E_{h}R_{h}>$ cannot be expected to reflect the $p_{t}$ behaviour,
 as they depend on several factors of the hadron cascade developement.
 
 To explain the enhancement in figure~\ref{fig:ervne}, an extremely 
 large increase in 
 primary mass, proportional to the primary energy, was introduced
 in the simulation \cite{NestRom} to transform 
 the general decrease
 of $<E_{h}R_{h}>$ versus $N_{e}$ in a visible increase. Such conjectures
 are no longer necessary and the agreement can be obtained without 
 a large increase of the primary mass.\\

 The $p_{t}$ dependence used here in the 3 options of CORSIKA covers
 a wide range of models; for instance, the values of $p_{t}$ in 
 option 2 are very close to the values used in SYBILL \cite{Gaisser}
 as shown in \cite{Knapp} by Knapp, Heck and Schatz, and we can conclude
 that the present measurements of hadrons at mountain altitudes 
 are in agreement
 with a majority of models used in EAS simulations. They require
 neither a large $<p_{t}>$ enhancement nor a noticeable increase
 of primary mass. Their significance will remain limited until the 
 resolution distance of the detectors is seriously improved.
 As it was advanced by Clay \cite{Clay},\lq\lq the  results in hadron 
 measurements depend critically on how well individual hadron signals
 are resolved in the hadron detector" and this has been confirmed
 by the present work.
 
 \ack{This work has been done under the contract
 of collaboration between IN2P3 Paris and IPJ Warsaw and the authors
 are indebted to both institutions for their support.\\
 JS expresses his gratitude to N.M.Nesterova and V.A.Romakhin
 for very valuable discussions about experimental details
 of the Tien Shan hadron calorimeter.}

 \end{document}